\documentclass[twocolumn,amsmath,amssymb,superscriptaddress,reprint,prl,aps]{revtex4-2}

\usepackage{upgreek}
\usepackage{relsize}
\usepackage{xcolor}
\usepackage{graphicx}
\usepackage{dcolumn}
\usepackage{bm}
\usepackage[colorlinks=true, citecolor=cyan, linkcolor=cyan,urlcolor=cyan]{hyperref}
\usepackage{todonotes}
\usepackage[dvipsnames]{xcolor}
\newcommand{\dir}{\bm{\mathcal{N}}} 

\usepackage[utf8]{inputenc}
\usepackage[T1]{fontenc}

\makeatletter
\@addtoreset{equation}{myequation}
\makeatother

\begin{document}
\newcommand{\ov}{\hat{\bm{n}}} 
\newcommand{\pv}{\bm{x}} 
\newcommand{\fv}{\hat{\bm{h}}} 
\newcommand{\la}{\mathcal{L}} 
\newcommand{\per}{\mathrm{Pe}_{R}} 
\newcommand{\tpa}{\tau_{\parallel}} 
\newcommand{\tpe}{\tau_{\perp}} 
\newcommand{\ef}{\mathfrak{f}} 
\newcommand{\ed}{\mathfrak{d}} 
\newcommand{\ecp}{\mathfrak{c}} 
\newcommand{\noi}{\zeta} 
\newcommand{\difc}{\mathcal{D}} 
\newcommand{\rop}{\bm{\mathcal{R}}_{\hat{\bm{n}}}} 
\newcommand{\lfp}{\ell_{\parallel}} %
\newcommand{\av}[1]{\langle #1\rangle} 

\newcommand{\warning}[1]{\todo[color=red, inline]{#1}}
\newcommand{\commAK}[1]{\todo[color=SeaGreen,size=\footnotesize]{#1}}
\newcommand{\commAKl}[1]{\todo[color=SeaGreen,inline]{#1}}
\newcommand{\ak}[1]{{\color{SeaGreen}#1}}
\newcommand{\commVS}[1]{\todo[color=magenta,size=\footnotesize]{#1}}
\newcommand{\commVSl}[1]{\todo[color=magenta,inline]{#1}}
\newcommand{\vs}[1]{{\color{magenta}#1}}

\title{Drift-diffusion interplay in active Brownian particles under orienting field}

\author{Andrey A. Kuznetsov}
\email{andrey.kuznetsov@univie.ac.at}
\affiliation{Computational and Soft Matter Physics, University of Vienna, Kolingasse 14-16, 1090 Vienna, Austria}
\author{Vittoria Sposini}%
\affiliation{Dipartimento di Fisica e Astronomia `G.~Galilei' -- DFA, Sezione INFN, Universit{\`a} di Padova, Via Marzolo 8, 35131 Padova, Italy
}%
\author{Sofia S. Kantorovich}
\affiliation{Computational and Soft Matter Physics, University of Vienna, Kolingasse 14-16, 1090 Vienna, Austria}
\author{Aleksei V. Chechkin}
\affiliation{Max Planck Institute of Microstructure Physics, Weinberg 2, 06120 Halle, Germany}
\affiliation{Faculty of Pure and Applied Mathematics
\\ Wroc{\l}aw University of Science and Technology, Wybrzeze Wyspianskiego 27,50-370 Wroc{\l}aw, Poland}
\affiliation{Akhiezer Institute for Theoretical Physics
National Science Center ``Kharkiv Institute of Physics and Technology'',\\
Akademichna Str. 1, 61108 Kharkiv, Ukraine}

\begin{abstract}
Magnetic active particles offer a versatile route to externally controlled microscale transport by combining self-propulsion with field-tunable orientation, as realized in both synthetic and living magnetic microswimmers. 
Here, we develop a theoretical framework for three-dimensional active Brownian motion in a uniform magnetic field, incorporating coupled translational and rotational dynamics and providing analytical approximations for low-order displacement moments. 
At long times, the
system dynamics reduces to a combination of enhanced diffusion and permanent drift absent in regular active Brownian particles. The field acts as an external controller, channeling activity toward one of these two types of motion.  At intermediate time scales, the interplay between rotational noise, self-propulsion, and magnetic alignment results in pronounced non-Gaussian displacement statistics. First-passage properties exhibit strong field sensitivity, highlighting the potential of magnetic guidance to optimize search processes and targeted delivery in active matter systems.
Theoretical predictions are validated by numerical simulations.
\end{abstract}

\maketitle


Active matter systems are composed of self-driven agents capable of continuously consuming energy to generate motion~\cite{Ramaswamy2010}.  
These systems operate far from thermodynamic equilibrium and exhibit rich collective behavior, including pattern formation, spontaneous flows, and large-scale organization~\cite{Cates2012,Marchetti2013,bechinger2016active}.
Their study offers opportunities to uncover new physical principles and inspire novel strategies for designing 
smart devices, 
functional materials, and  biomedical technologies.
As a result, active matter in recent years has attracted increasing attention across a wide range of disciplines~\cite{Gompper2025}.

Many living active systems demonstrate the ability to exploit the Earth's magnetic field for navigation -- from migratory animals such as birds and fishes~\cite{Hore2016,Mouritsen2018} to microorganisms such as magnetotactic bacteria~\cite{Blakemore1982,erglis2007dynamics,KLUMPP20191,Birjukovs2025}.
While the underlying physical mechanisms are different, ranging from complex biological sensing in higher organisms to simple magnetic alignment at the microscale, they all illustrate the fundamental role of the magnetic field in controlling active transport.
Inspired by these natural mechanisms, magnetic fields have also been used for the manipulation of synthetic microswimmers in the laboratory setting~\cite{Ju2025}.
In this context, it is important to differentiate between magnetically-driven swimmers, in which the field \textit{causes} the motion~\cite{Dreyfus2005,Tierno2008,Snezhko2009,Li2017,Han2020}, and magnetically-guided ones,
in which the field imposes an orientational bias and \textit{directs} an otherwise random self-propulsion~\cite{Baraban2012,Demirors2018,FernandezRodriguez2020,Wu2022,SCHYCK2025,Megha2026}.
The latter is the focus of the present work.

The most widely used mathematical model of the microscale active motion is the active Brownian particle (ABP) model~\cite{tenHagen2011,Romanczuk2012}.
Regardless of the propulsion mechanism, the model considers particles that move with a constant speed while their orientation undergoes rotational diffusion.
Despite being a minimal model, ABPs  constitute a powerful framework for understanding the out-of-equilibrium dynamics and is capable of capturing a wide variety of phenomena~\cite{zottl2023modeling}.
Unsurprisingly, the model has seen a lot of use in studies of magnetic/dipolar microswimmers.
Much of the literature focuses on 
emergent behavior in interacting swimmers -- 
both with~\cite{Romensky2015,Koessel2020,parage2025modulation,Telezki2025,Royall2025,Othman2025,Compagnie2026} and without~\cite{GuzmanLastra2016,Liao2020,Sansa2022,Kelidou2024,rosenberg2025windmilling,Musacchio2026} 
the orienting field. 
The field control of single-particle dynamics is much less explored and is typically studied by means of computer simulations~\cite{Codutti2019,KAISER2020}.
The analytical progress remains limited.

Perhaps, the most notable attempt to construct a theory of orientationally-biased ABPs was performed in Ref.~\cite{Takatori2014}, which predicted that the field can affect both the particles' mean velocity and effective diffusivity, with the latter becoming anisotropic. 
However, that theory was applicable only at times much longer than the ABP reorientation time. 
Another milestone work, Ref.~\cite{VidalUrquiza2017}, sought to address this constraint by introducing a time-dependent active diffusivity. 
Yet, Ref.~\cite{VidalUrquiza2017} conclusion that the motion of magnetic ABPs always transitions from ballistic to diffusive is at odds with recent experimental findings of Ref.~\cite{Megha2026}, which showed that the in-field mean squared displacement (MSD) of magnetic microswimmers remains ballistic at long times.
This makes it clear that active diffusivity alone is insufficient to fully explain the dynamics of these systems across the entire time range. 

Beyond diffusivity and the MSD, activity is known to affect other important properties of Brownian motion, including non-Gaussianity~\cite{Sevilla2015,Lemaitre2023,Baouche2024} and first-passage statistics~\cite{Basu2018,Baouche2025}, that are relevant for drug delivery applications~\cite{NEWBY201864,Singh2020,Lamirande2024}. 
However, the influence of the orienting field on these properties is not understood,
despite a growing interest in 
using magnetic swimmers for drug targeting~\cite{Felfoul2016,Kim2025}.

In this paper, we present a general theory of active Brownian motion in an orienting field, that overcomes the limitations of earlier studies and offers several new insights into the transport of magnetic ABPs in a broad range of propulsion speeds, field strengths, and time scales. 
The accuracy of theoretical predictions is confirmed by 
direct comparison with numerical simulations.

\section*{Model}

\begin{figure} 
\includegraphics[scale=1.15]{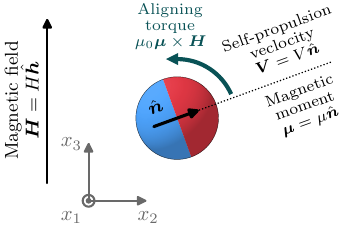}
\caption{\textbf{Model}. A 3D active Brownian particle is subjected to a homogeneous external field. The field induces an aligning torque and controls the direction of the particle self-propulsion by suppressing its rotational fluctuations. 
}
\label{fig:model}
\end{figure}

We consider an ABP of diameter $\sigma$,
which self-propels with velocity $\bm{V} = V \hat{\bm{n}}$~(Fig.~\ref{fig:model}). 
The orientation unit vector $\ov$~is 
fixed within the particle body frame, 
$V$ is constant.
The ABP moves in a three-dimensional viscous medium at temperature $T$,
being subjected to both translational and rotational Brownian motion.
The particle also possesses a permanent magnetic moment $\bm{\mu}$.
Similar to $\ov$, $\bm{\mu}$ is ``frozen'' in the particle body-fixed frame. 
Here we will assume that the two vectors are parallel {\it i.e.},~$\bm{\mu} = \mu \hat{\bm{n}}$.
This assumption is well justified for a number of real-life systems~\cite{Meng2018}, and serves as a convenient starting point for developing a theoretical framework~\cite{Takatori2014,VidalUrquiza2017}.
It should be emphasized, however, that for certain types of synthetic swimmers the directions of $\bm{\mu}$ and $\bm{V}$ do not coincide~\cite{KAISER2020,Megha2026}. 
This scenario is left for future studies. 
The system is subjected to a homogeneous magnetic field $\bm{{H}} = H\hat{\bm{h}}$ oriented along the $z$-axis, \textit{i.e.}, $\fv = [0,0,1]$.
The orientational coupling between the ABP and the field is governed by the  Zeeman potential $E_H = - \mu_0 \, \bm{\mu} \cdot \bm{{H}}$. 
Translational and orientational overdamped motion of such ABP are described by Langevin equations~\cite{zottl2023modeling,coffey2004langevin}

\begin{subequations} \label{eq:lang}
\begin{gather} 
    \frac{d \bm{r}}{dt} = V\bm{\hat{n}} + \sqrt{2 D_T} \bm{\noi}_T, \label{eq:lang_t}\\
    \frac{d\hat{\bm{n}}}{dt} = \bm{\omega} \times \hat{\bm{n}}, \:\:\:
    \bm{\omega}  = \frac{\mu_0}{\gamma_R} \left[\bm{\mu} \times \bm{H}\right] + \sqrt{\frac{1}{\mathcal{T}_R}} \bm{\noi}_R. \label{eq:lang_r}
\end{gather}
\end{subequations} 

\noindent Here, $\bm{r}$ is the particle position, 
$\bm{\omega}$ is its angular velocity,
$D_T = {k_B T}/{\gamma_T}$ is the translational diffusion coefficient,
$\mathcal{T}_R = \gamma_R / 2 k_B T$ is the rotational diffusion time,
$k_B$ is the Boltzmann constant,
$\gamma_T$ and $\gamma_R$ are the translational and rotational friction coefficients, respectively,
$\bm{\noi}_T(t)$ and $\bm{\noi}_R(t)$ are noise vectors with zero mean and autocorrelation functions $\langle \noi_{T/R,i}(t_1) \noi_{T/R,j}(t_2)\rangle = \delta_{ij} \delta(t_1 - t_2)$.
The first term on the r.h.s. of $\bm{\omega}$ in Eq.~(\ref{eq:lang_r}) corresponds to the \textit{magnetic torque}.
{The field does not cause any \textit{magnetic forces} --
neither the Lorentz force, which would appear if the ABP were charged~\cite{Vuijk2020}, nor the Kelvin force, which would require a field gradient~\cite{Butcher2023}. 
Here, the field affects the translational motion exclusively via velocity alignment.}

Henceforth, we will operate with a non-dimensional variant of the model, 
using the diameter~$\sigma$ as a {unit of distance} 
and the {rotational} relaxation time~$\mathcal{T}_R$ as a {unit of time}. 
The dimensionless position vector and time are
$\bm{x} = {\bm{r}}/{\sigma} = [x_1, x_2, x_3]$ and $\tau = {t}/{\mathcal{T}_R}$,
respectively.
Upon non-dimensionalization, Eqs.~(\ref{eq:lang}) take form

\begin{subequations} \label{eq:lang_nd}
\begin{gather}  
    \frac{d\pv}{d \tau} = \per \:{\ov} + \sqrt{2 \difc}\:\bm{\mathcal{\zeta}}^*_T(\tau), \\
    \frac{d\ov}{d \tau} = \frac{\xi}{2}\:\left[\hat{\bm{n}} \times \hat{\bm{h}}\right]\times \hat{\bm{n}} + \bm{\mathcal{\zeta}}^*_R(\tau) \times \hat{\bm{n}}. \label{eq:lang_r_nd}
\end{gather}
\end{subequations}

\noindent Here, $\bm{\mathcal{\zeta}}^*_T(\tau) = \bm{\zeta}_T(t) \sqrt{\mathcal{T}_R}$ and $\bm{\mathcal{\zeta}}^*_R(\tau) = \bm{\zeta}_R(t) \sqrt{\mathcal{T}_R}$ are dimensionless noise vectors,
and the three main parameters of the model are defined as

\begin{equation} \label{eq:parameters}
    \xi = \frac{\mu_0 \mu H}{k_B T}, \:\:\:\: \per = \frac{\mathcal{T}_R V}{\sigma}, \:\:\:\: \difc =\frac{\mathcal{T}_R  D_T}{\sigma^2}. 
\end{equation}

\noindent 
The dimensionless field strength, $\xi$, characterizes the field ability to suppress rotational fluctuations and to orient the ABP along $\hat{\bm{h}}$.  
In ferrofluid physics and nanomagnetism, it is known as the {\it Langevin parameter}~\cite{raikher1994effective}.
The {\it rotational P\'eclet number}, $\per$, is a measure of how far the ABP can travel in a given direction before the rotational noise knocks it off course~\cite{zottl2023modeling}. 
Alternatively, one can use the translational  P\'eclet number
$\mathrm{Pe} = {V \sigma}/{D_T}$~\cite{kurzthaler2016intermediate}. 
The two are connected as $\per = \difc \mathrm{Pe}$, 
where $\difc$ is the ratio of rotational and translational diffusive times, 
which acts here as the
\textit{dimensionless diffusion coefficient}.
For a sphere at low Reynolds numbers in a medium of viscosity $\eta$, 
$\gamma_T = 3 \pi \eta \sigma$ and $\gamma_R = \pi \eta \sigma^3$, so $\difc = 1/6$~\cite{happel1983low}. 
We use this value for all numerical evaluations.

Chances of finding the ABP in a given state are described by the probability density function (PDF) $W = W(\bm{x},\hat{\bm{n}},\tau)$. 
{Interpreting Eq.~(\ref{eq:lang_nd}) in the Stratonovich sense,} we can formulate
the Fokker-Planck equation (FPE) governing the PDF evolution as~\cite{risken1989,garcia1998langevin}

\begin{multline} \label{eq:fpe}
\frac{\partial W}{\partial \tau} = - \per \left(\hat{\bm{n}} \cdot \bm{\nabla} \right) W + \difc \bm{\nabla}^2 W \\ - \frac{\xi}{2}\rop \cdot \left( [\hat{\bm{n}} \times \hat{\bm{h}}]W\right) + \frac{1}{2}\rop^2 W,
\end{multline}

\noindent 
where $\rop \equiv \hat{\bm{n}} \times \partial/\partial \hat{\bm{n}}$ is the infinitesimal rotation operator.
It is convenient to  
introduce the orientation-averaged and space-averaged marginal PDFs, respectively,

\begin{equation} \label{eq:marginal}
P(\pv,\tau) = \int W d \ov, \:\:\:\: Q(\ov,\tau) = \int W d \pv.  
\end{equation}

\noindent 
The latter, $Q$, describes the chances of the ABP to have a given orientation regardless of its positioning. 
The field is uniform in space, and the rotation is not affected by the translational motion.
Thus, one should expect that the magnetic moment will eventually reach an equilibrium state described by the Boltzmann distribution 

\begin{equation} \label{eq:q_eq}
Q_\mathrm{eq}(\ov) = \frac{\exp(-E_H/k_B T)}{\int \exp(-E_H / k_B T) d \ov} = \frac{\xi}{4\pi \sinh \xi} \exp(\xi  \hat{\bm{h}}\cdot\ov) .
\end{equation}

\noindent
For simplicity, we use this equilibrium distribution as the initial state for the orientation and consider the particles to start at the origin,
\textit{i.e}., the 
initial condition is $W(\bm{x},\hat{\bm{n}},0) = \delta(\bm{x}) Q_\mathrm{eq}(\hat{\bm{n}})$.

{Here, we develop a solution scheme for Eq.~(\ref{eq:fpe}),
which allows one to calculate {the} PDF moments at arbitrary $\tau$, $\xi$ and $\per$.
{The} exact solutions, however, cannot be represented in a closed form.
So, we put forward
{a} second approach based on the \textit{effective field approximation} (EFA).
It
was origianlly developed to study orientational kinetics of ferrofluids~\cite{martsenyuk1974kinetics,raikher1994effective} and subsequently was adopted to a number of superparamagnetic systems~\cite{Rusakov2021,MUSIKHIN2023,Tripathi2025}.
We generalize it to account for ABPs translational degrees of freedom and use it to obtain approximate formulas for low-order PDF moments.
See \textit{Methods} for details.
}

\begin{figure*}
\includegraphics[scale=1]{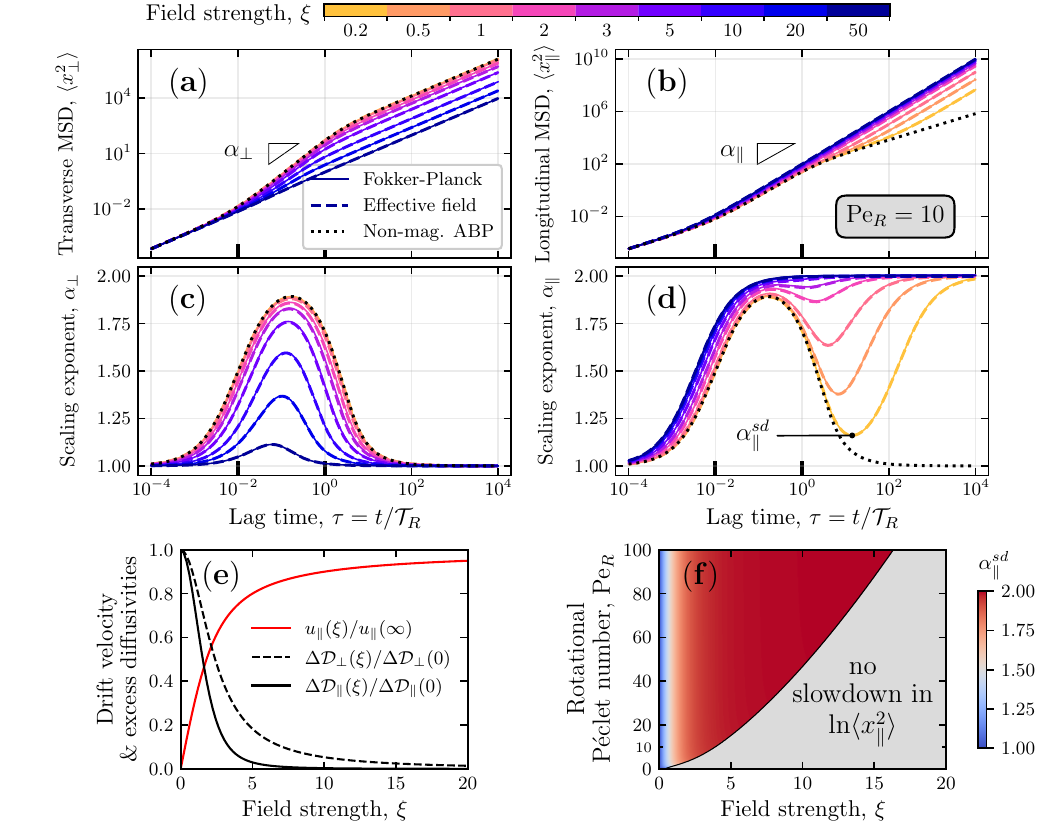}
\caption{\textbf{Field-controlled drift and anisotropic diffusion in the mean squared displacement (MSD)}. \textbf{(a)} MSD vs. lag time orthogonal to and \textbf{(b)} along to field, $\text{Pe}_R = 10$. 
Solid lines -- Fokker-Planck solutions for different fields $\xi$, 
dashed lines -- effective field approximation (EFA) [Eq.~(\ref{eq:msd})],
dotted lines -- non-magnetic active particles ($\xi = 0$).
Two large vertical ticks denote non-magnetic crossover times, $\tau^\mathrm{diff} = 6\difc/\per^2 = 10^{-2}$ and $\tau = 1$.
\textbf{(c), (d)} MSD scaling exponents 
$\alpha_{\rho} = {d \ln \langle x^2_{\rho} \rangle}/{d \ln \tau}$, $\rho = \: \perp$ {(c)} or  $\parallel$ {(d)}, as a function of lag time. Line styles
are the same as in (a), (b).
\textbf{(e)} Drift velocity and long-term excess diffusivities vs. field according to EFA. 
Colored line -- velocity [Eq.~(\ref{eq:drift})], solid black line -- longitudinal diffusivity, dashed black line -- transverse diffusivity [Eq.~(\ref{eq:delta_ds})]. Velocity is normalized by its value at $\xi \rightarrow \infty$, while diffusivities are normalized by zero-field values. 
\textbf{(f)} ``Slowdown'' exponent $\alpha^{sd}_{\parallel}$ at different $\text{Pe}_R$ and $\xi$ according to EFA. It is determined from $d \alpha_{\parallel} / d \tau = 0$, $d^2 \alpha_{\parallel} / d \tau^2 > 0$ [see Panel (d)]. Within the gray area, $\alpha_{\parallel}$ monotonically increases with time.
}
\label{fig:msd}
\end{figure*}

\section*{Results}

\paragraph{Mean and mean squared displacements.}
{In an unbounded domain, the field introduces an axial symmetry. 
So, it makes sense to separately consider the motion along the field {($x_{\parallel} \equiv x_3$)} and in the orthogonal plane {($x_{\perp} \equiv \sqrt{x_1^2 + x_2^2}$)}.
In the latter case, the symmetry requires zero mean displacements, \textit{i.e.}, ${\av{x_1} = \av{x_2} = 0}$. 
However, in the longitudinal direction, the velocity aligning effect of the field causes a \textit{permanent drift}. 
For an equilibrium initial orientation, the mean displacement is

\begin{equation} \label{eq:drift}
    \av{x_{\parallel}} = u_{\parallel} \tau, \:\:\:\:  u_{\parallel} = \la(\xi)\per.    
\end{equation}

\noindent
Here, $u_{\parallel} $ is the dimensionless drift velocity, and} $\la(\xi) = \coth \xi - 1/\xi$ is the Langevin function
describing the magnetization curve of an ideal superparamagnet:
without the field, $\xi = 0$, magnetic moments are oriented at random with zero mean ($\la(0) = 0$), while an infinitely strong field fully aligns them ($\la(\infty) = 1$)~\cite{shliomis1974magnetic}. 
For the second moments, the EFA produces closed-form expressions
\begin{subequations} \label{eq:msd}
\begin{gather}
    \left\langle x^2_{\parallel} \right\rangle = u^2_{\parallel} \tau^2 + 2 \Delta \difc_{\parallel}\left(\tau/\tpa + e^{-\tau/\tpa} - 1\right) \tpa + 2 \difc \tau, \label{eq:msdpa} \displaybreak[0]\\
    \left\langle x^2_{\perp} \right\rangle = 4\Delta \difc_{\perp}\left(\tau/\tpe + e^{-\tau/\tpe} - 1\right) \tpe + 4 \difc \tau. \label{eq:msdpe}
\end{gather}
\end{subequations}

\noindent
Here, $\tpa$ and $\tpe$ are the {relaxation times} of the magnetic moment along and orthogonal to the field~\cite{martsenyuk1974kinetics},  

\begin{equation}
    \tpa = \frac{\xi}{\la(\xi)} \frac{d \la(\xi)}{d \xi},  
    \:\:\:\: 
    \tpe = \frac{2 \la(\xi)}{\xi - \la(\xi)}, \label{eq:taus} 
\end{equation}

\noindent
$\Delta \difc_{\parallel}$ and $\Delta \difc_{\perp}$ are the 
long-time excess diffusivities, 

\begin{equation}
    \Delta \difc_{\parallel} = \per^2 \tpa \frac{d \la(\xi)}{d \xi}, 
    \:\:\:\: 
    \Delta \difc_{\perp} =  \per^2 \tpe \frac{\la(\xi)}{\xi}. \label{eq:delta_ds}
\end{equation}

\noindent
At ${\tau \rightarrow \infty}$, ${\langle  (x_{\parallel} - \av{x_{\parallel}})^2\rangle/ 2 \tau = \Delta \difc_{\parallel} + \difc}$ and ${ \langle x_{\perp}^2 \rangle/ 4 \tau = \Delta \difc_{\perp} + \difc}$.
Our predictions for 
$\Delta \difc_{\parallel,\perp}$ 
agree with
Refs.~\cite{Takatori2014,VidalUrquiza2017}
(see the EFA subsection of \textit{Methods} for more details).
However, Eqs.~(\ref{eq:msd}) expand upon these earlier works and provide a complete and compact representation of the second displacement moments across 
the entire time range.
Its accuracy is confirmed by a direct comparison with the FPE numerical solutions (see Figs.~\ref{fig:msd}~(a)--(d), solid and dashed lines -- for all times and field values, the overall relative error of the EFA remains below one percent for the longitudinal motion and below four percents for the transverse case).
Here, we will refer to $\av{x_{\parallel,\perp}^2}$ as the mean squared displacement (MSD), meaning the displacement from the initial position $\pv(0) = \bm{0}$.
In case of $\av{x_{\parallel}^2}$, it must not be confused with the variance $\av{\delta x^2_{\parallel}} = \av{x^2_{\parallel}} - \av{x_{\parallel}}^2$, \textit{i.e.}, the mean squared displacement from the current mean position ($\delta x_{\parallel} \equiv x_{\parallel} - \av{x_{\parallel}}$).

In the non-magnetic limit $\xi = 0$,  
$\tau_{\parallel} = \tau_{\perp} = 1$,
$u_{\parallel} = 0$, and 
Eqs.~(\ref{eq:msd}) reduce
to the known exact formula for the three-dimensional MSD of regular ABPs, ${\langle x^2 \rangle = \langle x^2_{\parallel} \rangle + \langle x^2_{\perp} \rangle = 2 \per^2(\tau + e^{-\tau} - 1) + 6\difc \tau}$~\cite{tenHagen2011,kurzthaler2016intermediate}. 
The corresponding curves are reported in Figs.~\ref{fig:msd}~(a)--(b) with dotted lines.
They demonstrate three distinct dynamic regimes. 
Initially, the displacement occurs mainly because of regular Brownian diffusion, and the MSD scaling exponent $\alpha$ ($\langle x^2 \rangle \propto \tau^\alpha$) is close to one (dotted lines in Fig.~\ref{fig:msd} ~ (c)--(d)).
However, the particle keeps propelling itself with constant velocity in whichever direction it was initially facing, and at intermediate times the motion acquires a persistent (superdiffusive) character with $\alpha > 1$. 
At $\per > 1$, the crossover time between these two regimes is $\tau^{\mathrm{diff}} \simeq 6 \difc/ \per^2$. 
Around $\tau \simeq 1$ a second crossover occurs -- the rotational noise starts to randomize the direction of the particle self-propulsion, and $\alpha$ starts to decrease. 
For ${\tau \gg 1}$, the ensemble-averaged motion becomes again diffusive, but with an enhanced diffusion coefficient ${\per^2/3 + \difc}$.   

Solid and dashed curves in Figs.~\ref{fig:msd}~(a)--(d) show how the MSD in a previously mentioned nonmagnetic case evolves once the magnetic field is applied. 
Figs.~\ref{fig:msd}~(a) and (c) represent the transverse dynamics.
The latter only quantitatively deviates from the non-magnetic case: the field favors the out-of-plane orientation of the velocity, gradually reducing the in-plane effect of self-propulsion.
The crossover time from the initial diffusive to the superdiffusive regime, $\tpe^{\mathrm{diff}}$, increases with the field.
At $\xi < 1$, it can be estimated from Eq.~(\ref{eq:msdpe}) as ${\tpe^\mathrm{diff} \simeq \tau^\mathrm{diff}(1 + \xi^2/15)} $. 
On the other hand, the relaxation  time, which heralds the loss of persistence, decreases: ${\tpe \simeq 1 - \xi^2/10}$. 
The excess diffusivity $\Delta \difc_{\perp}$ also drops, approaching zero for $\xi \gg 1$ (Fig.~\ref{fig:msd}~(e), dashed line).
At infinitely large fields, the transverse motion is purely passive.

The longitudinal motion, Figs.~\ref{fig:msd}~(b)~and~(d),
differs qualitatively from the non-magnetic case, 
being defined by the interplay between drift [the first term in Eq.~(\ref{eq:msdpa})] and excess diffusivity [the second term].
Both effects show the same $\per^2$ scaling with activity, 
but have opposite field trends (Fig.~\ref{fig:msd}~(e), solid lines). 
Thus, the field acts as a control parameter allowing to channel the activity towards one of these two types of motion.
The activity manifests itself only as diffusion at $\xi = 0$ and only as drift at $\xi \gg 1$.

Contrary to non-magnetic and transverse MSD curves, 
that exhibit three dynamic regimes,
$\av{x_{\parallel}^2}$ shows either four or two, 
depending on $\per$ and $\xi$.
The first scenario is realized when the field is not too large to fully suppress the ABP orientational fluctuations.
It is well illustrated by the $\xi = 0.2$ curve in Fig.~\ref{fig:msd}~(d).
The crossover from the first (diffusive) to the second (persistent) regime happens at ${\tpa^{\mathrm{diff}} \simeq \tau^\mathrm{diff}(1 - \xi)}$.
Near $\tpa \simeq 1 - 2\xi^2/15$, 
the growth of the MSD slows down due to orientational fluctuations of the velocity vector and 
$\alpha_{\parallel}$ starts decreasing (third regime).
However, because of the weak drift, net displacements along the field accumulate. Eventually, the exponent starts increasing again towards the ballistic value $\alpha_{\parallel} = 2$ (fourth regime).
In the third regime,
$\alpha_{\parallel}(\tau)$ has a local minimum, which in Fig.~\ref{fig:msd} (d) is designated as the ``slowdown'' exponent $\alpha^{sd}_{\parallel}$.
Fig.~\ref{fig:msd} (f) gives values of $\alpha^{sd}_{\parallel}$ at different fields and P{\'e}clet numbers. 
In the gray area of Fig.~\ref{fig:msd}~(f), a different scenario with only two dynamical regimes takes place.
Here, the field aligning effect is strong, so the motion never looses its persistence at $\tau > \tpa^\mathrm{diff}$ and the self-propulsion direction does not deviate from~$\fv$. 

Our theoretical prediction of an altering number of dynamical regimes and a longitudinal ``slowdown''
is in good qualitative agreement with experimental results recently reported in Ref.~\cite{Megha2026},
where the trajectories of hematite Janus microswimmers of varying magnetic strength were studied in the geomagnetic field.
In these experiments, swimmers with higher magnetic moments (\textit{i.e.}, higher~$\xi$) demonstrated straight trajectories with consistently ballistic MSD, 
whereas weakly magnetic particles showed meandering trajectories accompanied by a decrease in the MSD exponent at intermediate time scales
(compare our Figs.~\ref{fig:msd}~(b) and (d) with Fig. 4 of Ref.~\cite{Megha2026}). 


\begin{figure*}
\includegraphics[width=\textwidth]{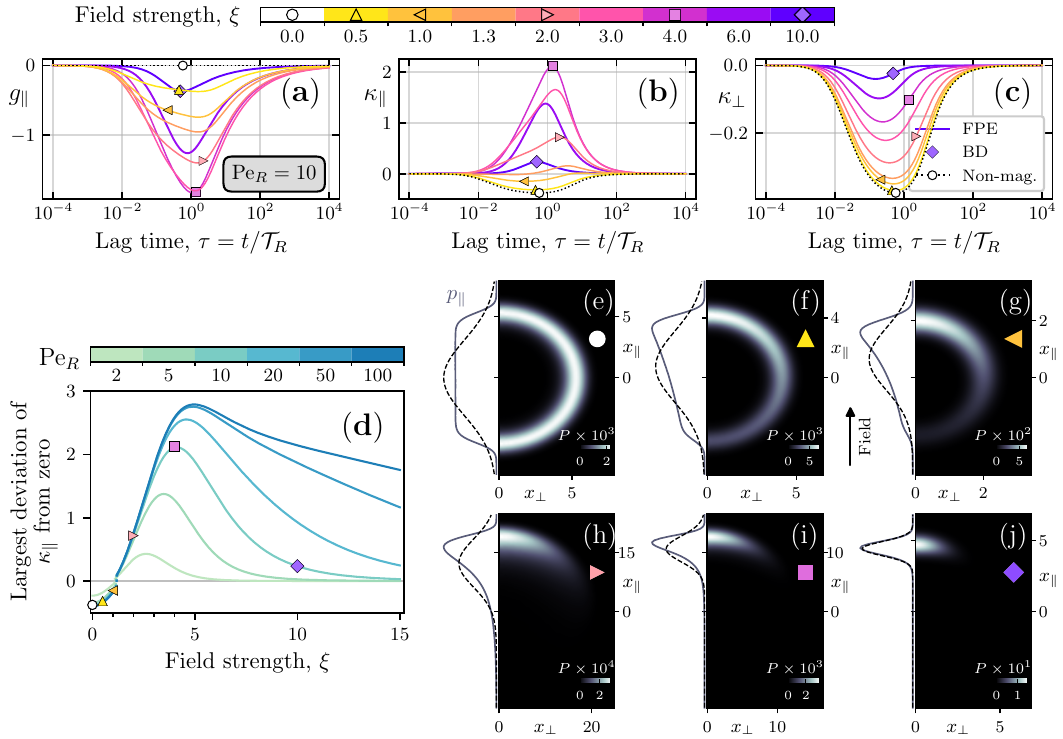}
\caption{\textbf{Deviation from Gaussianity at intermediate time scales}. \textbf{(a)} Longitudinal skewness $g_{\parallel}$, \textbf{(b)} longitudinal non-Gaussianity parameter (NGP) $\kappa_{\parallel}$ and \textbf{(c)} transverse NGP $\kappa_{\perp}$ vs. lag time, $\text{Pe}_R = 10$. 
Solid lines -- Fokker-Planck (FPE) solutions for different fields~$\xi$, 
dotted lines -- FPE for non-magnetic ABP ($\xi = 0$), symbols -- Brownian dynamics (BD). 
\textbf{(d)}~Largest deviations of $\kappa_{\parallel}$ from zero vs. field. Solid lines -- FPE for different $\per$, symbols -- BD for $\per = 10$.
\textbf{(e)}--\textbf{(j)} Particle spatial distributions at the moment, when $\kappa_{\parallel}$ deviates from zero the most. BD results for $\text{Pe}_R = 10$ and different fields: $\xi = 0$ (e), 0.5 (f), 1 (g), 2 (h), 4 (i) and 10 (j). Panels show orientation-averaged PDFs $P = P(x_{\parallel}, x_{\perp}, \tau)$. 
Solid lines to the left of panels are longitudinal marginal distributions $p_{\parallel} = 2\pi \int x_{\perp} P d x_{\perp}$, 
dashed lines are Gaussian effective field predictions for $p_{\parallel}$.}
\label{fig:ngp}
\end{figure*}

\paragraph{When is the Gaussian description sufficient?}

{The approximation obtained in} Eq.~(\ref{eq:msd}) was derived under the assumption that the ABP spatial distribution is Gaussian.
As shown in the previous section, this is still sufficient to accurately estimate the first two displacement moments. 
However, the ABP dynamics is known to be generally non-Gaussian~\cite{kurzthaler2016intermediate}.
Therefore, it is instructive to look at the non-Gaussian features of our model and to understand how are they affected by the field. 
We make use of the FPE numerical solutions and focus on two characteristics -- skewness and non-Gaussianity parameter (NGP)~\cite{tenHagen2011,Hofling2013}.
The skewness $g_{\parallel}$ is only calculated in the longitudinal direction,
while the NGP $\kappa_{\parallel,\,\perp}$ is calculated both along and perpendicular to the field: 

\begin{equation}
    g_{\parallel} = \frac{\langle \delta x_{\parallel}^3 \rangle}{
    \langle \delta x_{\parallel}^2\rangle^{3/2} 
    }, \:\:\: \kappa_{\parallel} = \frac{1}{3}\frac{\langle \delta x_{\parallel}^4\rangle}{\langle \delta x_{\parallel}^2\rangle^2} - 1, \:\:\: \kappa_{\perp} = \frac{1}{2}\frac{\langle  x_{\perp}^4\rangle}{\langle x_{\perp}^2\rangle^2} - 1.
\end{equation}

\noindent 
Dotted lines in Figs.~\ref{fig:ngp}~(a)--(c) show time dependencies of these parameters for non-magnetic ABPs. 
The skewness remains zero, since at $\xi =0$ the problem is isotropic and there is no asymmetry in the ABPs spreading pattern (assuming equilibrium initial conditions).
The NGP instead shows three regimes, similar to the MSD. At both small ($\tau \ll \tau^{\mathrm{diff}}$) and large times ($\tau \gg 1$), when the motion is diffusive, the NGP is close to zero, indicating the Gaussian character of the PDF. 
At $\tau \sim 1$, within the superdiffusive time window, 
particles can self-propel from the initial position in any direction with equal probability.
As a result, the PDF in this regime resembles a spherical shell of radius $\sim \per \tau$, while its projection on any direction is flat with heavy shoulders and short tails (see Fig.~\ref{fig:ngp}~(e)).
Thus, the NGP is negative and approaches the uniform distribution value ${\kappa = - 0.4}$~\cite{DeCarlo1997}.

At non-zero fields, one thing remains true -- the PDF always starts from a Gaussian shape and always returns to it at $\tau \rightarrow \infty$.
Deviations from Gaussianity only take place at the intermediate time scale.
For the transverse motion, as with the MSD, the field does not cause qualitative changes, but gradually reduces the effect of activity -- the window of non-Gaussian behavior narrows, while the maximum of $|\kappa_{\perp}|$ decreases (see Fig.~\ref{fig:ngp}~(c) and compare it to Fig.~\ref{fig:msd}~(c)).
The longitudinal motion is more intricate. 
Both $g_{\parallel}$ and $\kappa_{\parallel}$ show non-monotonic field dependencies. 
The skewness is non-positive throughout the investigated P{\'e}clet range ($\per \le 100$), $-3 < g_{\parallel} \le 0$.
The largest deviations from zero occur at $\xi \sim 5$, while at stronger fields the skewness vanishes. 
The longitudinal NGP starts with negative values but changes sign at $\xi \sim 1$ [Figs.~\ref{fig:ngp}~(d)].
Thus, the model demonstrates a \textit{field-controlled transition from platykurtic to leptokurtic behavior}~\cite{DeCarlo1997}.
As with $g_{\parallel}$, the NGP takes the highest value at $\xi \sim 4-5$ while for larger fields it decreases.
Some intuition for this non-monotonic behavior could be gained from 
{Brownian dynamics} (BD) snapshots in Figs.~\ref{fig:ngp} (e)-(j) (see \textit{Methods} for BD details).
Snapshots show how the PDF shape at intermediate times evolves with increasing field.
The initial orientational distribution of particles changes with the field and so does the pattern of their self-propulsion -- it goes from a spherical shell into a contracting spherical cap (crescent).
The two shoulders of the longitudinal marginal distribution $p_{\parallel}(x_{\parallel}, \tau) = 2\pi \int x_{\perp} P d x_{\perp}$ evolve asymmetrically (hence, a non-zero skewness).  
The shoulder in front (with respect to the field direction) 
becomes more pronounced and eventually develops into a sharp peak. 
The opposite shoulder,  
which corresponds to the particles moving against the field,
flattens and turns into a long heavy tail at $\xi \gtrsim 1$ (compare Fig.~\ref{fig:ngp}~(g), for which  $\xi = 1$ and $\kappa_{\parallel} < 0$, to Fig.~\ref{fig:ngp}~(h), for which  $\xi = 2$ and $\kappa_{\parallel} > 0$).
At $\xi \gg 1$, particles propel along the field and only deviate from the mean position due to regular Brownian motion, which ensures vanishing NGP at large fields.

\begin{figure*}[t]
\includegraphics[scale=1]{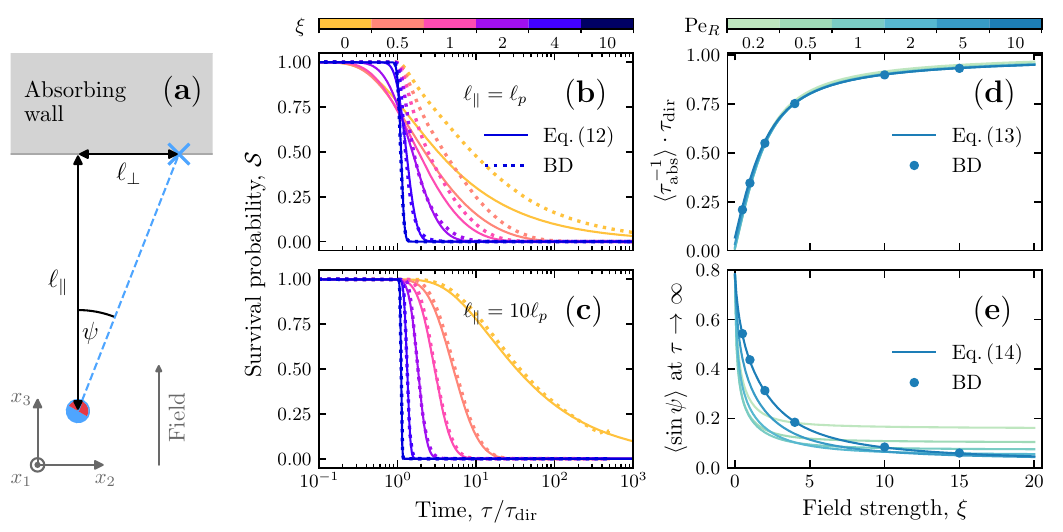}
\caption{\textbf{First-passage statistics: trade-off between absorption rate and surface coverage}. 
\textbf{(a)} Schematic illustration of the first-passage problem in a semi-infinite domain.
\textbf{(b)} Survival probability vs. time for the initial  distance $\ell_{\parallel} = \ell_{p} $ and \textbf{(c)} $\ell_{\parallel} = 10 \ell_{p}$. 
For both, $\text{Pe}_R = 10$.
Solid lines -- Eq.~(\ref{eq:surv}), 
dots -- Brownian dynamics (BD).
\textbf{(d)} Mean inverse absorption time and \textbf{(e)} mean sine of the spread angle vs. field strength at $\ell_{\parallel} = 100$. 
Symbols -- BD for $\text{Pe}_R = 10$.
Solid lines -- theory for different~$\text{Pe}_R$,
Eq.~(\ref{eq:inv_time}) (d) and Eq.~(\ref{eq:sin}) (e).
}
\label{fig:fps}
\end{figure*}

\paragraph{First-passage statistics.}

Finally, let us give an example of how the obtained results could be applied to more practical problems, such as magnetic drug targeting and cargo transport.
Assume that the ABP is guided by the field towards a stationary target initially positioned at a distance $\lfp$.
The target is much larger than the particle and thus can be approximated by an infinite wall orthogonal to the field (Fig.~\ref{fig:fps}~(a)).
One is interested in how long it takes for the particle to reach the target on average. 
This is equivalent to a standard first-passage problem in a semi-infinite domain with an absorbing boundary~\cite{Redner2001}. 

We now know that at sufficiently long times ${\tau \gg 1}$ the particle PDF reduces to a drifting Gaussian distribution with velocity $u_{\parallel}$ and constant diffusion coefficients $\difc^\text{eff}_{\parallel, \perp} = \Delta \difc_{\parallel, \perp} + \difc$.
For such distribution, the {survival probability} in a semi-infinite domain is  

\begin{equation} \label{eq:surv}
    \mathcal{S}(\tau) = \frac{1}{2} \Bigg[ \mathrm{erfc} \left( \frac{u_{\parallel} \tau - \ell_{\parallel}}{2 \sqrt{\difc^\text{eff}_{\parallel}\tau}}\right) - w \: \mathrm{erfc} \left( \frac{ u_{\parallel} \tau + \ell_{\parallel}}{2 \sqrt{\difc^\text{eff}_{\parallel}\tau}}\right) \Bigg],
\end{equation}

\noindent
where $w = \exp \Big(l_{\parallel} u_{\parallel}/\difc^\text{eff}_{\parallel}\Big)$ 
(derivations of all the formulas from this section are provided in \textit{Methods}). 
$\mathcal{S}$ shows how likely it is that the particle did not yet reach the target by time $\tau$.
One can expect Eq.~(\ref{eq:surv}) to work for ABPs if the initial distance is large, which gives the PDF enough time to acquire its long-term Gaussian form.
Quantitatively it means that ${\ell_{\parallel} \gg \ell_p}$, 
where $\ell_p$ is the persistence length, \textit{i.e.}, 
the distance ABP covers before the noise reorients it.
In our dimensionless units, $\ell_p = \per$.
A direct comparison of Eq.~(\ref{eq:surv}) to BD simulations is shown in Figs.~\ref{fig:fps}~(b) and (c). 
It is clear that for non-magnetic ABPs close to the wall (${\xi = 0}$, ${\ell_{\parallel} = \ell_p}$) theory indeed overestimates the rate of the particle absorption (Fig.~\ref{fig:fps}~(b)). 
At $\lfp > \ell_p$, the agreement markedly improves (Fig.~\ref{fig:fps}~(c)).
The accuracy of the formula further increases with the field -- at $\xi = 10$, the predictions are good even for a closely positioned target.
At $\xi \gg 1$ and $\per \gg 1$, the particle 
swims directly towards the wall and reaches it in $\tau_\text{dir} = \lfp/\per = \lfp/\ell_p$.
This time is a lower bound for the mean absorption time $\av{\tau_\text{abs}}$ (or the \textit{mean first-passage time}).
Since for Brownian particles without drift the first-passage time is divergent~\cite{Redner2001}, we instead can use the inverse time $\av{\tau_\text{abs}^{-1}}$ to estimate the  particle absorption rate. 
Eq.~(\ref{eq:surv}) corresponds to

\begin{equation} \label{eq:inv_time}
    \av{\tau_\text{abs}^{-1}} = \frac{2\difc^\text{eff}_{\parallel} }{\lfp^2} + \frac{ u_{\parallel}}{\lfp}.
\end{equation}

\noindent
At large initial distances (equivalently, at $\tau_\text{dir} \gg 1$) and $\per \ge 1$,
the formula reduces to $\av{\tau_\text{abs}^{-1}}\, = \tau_\text{dir}^{-1} \la(\xi) + \mathcal{O}(\tau_\text{dir}^{-2})$. 
Thus, the  normalized absorption rate nonlinearly grows with the field, closely following the Langevin function (Fig.~\ref{fig:fps} (d)).


Beyond the mean first-passage time, another potentially relevant quantity is the spatial distribution of particles over the target surface upon arrival. 
In targeted drug delivery, for instance, the ability to control the surface coverage of the target could improve the spatial distribution of drug release as well as limit off-target effects~\cite{blanco2015principles}. 
A rigorous formulation of this problem for magnetic microswimmers is left for future work. 
Nevertheless,  
our simple setup already provides useful insights into the interplay between arrival times and surface coverage.
Let us denote the distance between the target center and the point of contact as $\ell_{\perp}$. 
For non-drifting Brownian particles, $\av{\ell_{\perp}} \rightarrow \infty$ as $\tau \rightarrow \infty$. 
So, we look at the ``spread angle'' $\psi$, 
which remains finite regardless of the drift velocity (${0 \le \psi < \pi /2}$).
At $\tau \rightarrow \infty$,

\begin{equation} \label{eq:sin}
    \av{\sin \psi} = \nu \int_1^{\infty} \frac{w^\frac{1-y}{2}}{\sqrt{y^2 - 1}\:(\nu + y^2 - 1)^{3/2}} d y, 
\end{equation}

\noindent
where  $\sin \psi = 1/\sqrt{1 + \lfp^2/\ell_{\perp}^2}$,
${\nu = \difc^\text{eff}_{\parallel}/\difc^\text{eff}_{\perp}}$.
At $\xi = 0$, $\nu = 1$, $w = 1$
and $\av{\sin \psi} = \pi / 4 \simeq 0.8$.
With increasing~$\xi$, the mean 
sine quickly decreases (Fig.~\ref{fig:fps} (e)).

\section*{Conclusions}

In this work, we have theoretically investigated 
the effect of a uniform velocity-aligning field
on the dynamics of non-interacting ABPs.
One of the key results is the approximation Eq.~(\ref{eq:msd}) for the second translational moment of particles along and orthogonal to the field.
It demonstrates that the field-induced drift can modify the time dependence of the MSD and change the number of observed dynamical regimes. 
This result provides an explanation for recent experimental findings of Ref.~\cite{Megha2026}, where qualitatively different transport dynamics were reported for magnetic microswimmers with different intrinsic magnetic strengths.

The field-controlled interplay of drift and diffusion is also shown to be a major factor in the ABP first-passage statistics.   
While the drift largely speeds up the absorption [Eq.~(\ref{eq:inv_time})], 
the reduction in effective diffusivity limits the surface coverage of the target [Eq.~(\ref{eq:sin})]. 
Therefore simply increasing the field might not be an optimal solution for a given cargo transport application and an additional parametric fine-tuning might be required. 
Note that our treatment of the first-passage problem was restricted to a semi-infinite domain with large initial distances to the target. 
As a prospective direction, it would be interesting to investigate more realistic geometries and initial distances comparable to the ABP persistence length. 
In the latter case, we expect that the dependency of the PDF shape on field (Fig.~\ref{fig:ngp}) will play a prominent role.

The analytical progress achieved in this work was made possible by the use of the classical effective field method~\cite{raikher1994effective}, 
which we generalized to account for the coupling between ABP translational and rotational degrees of freedom. 
Although we have focused here exclusively on the static field dynamics, 
the method allows one to explore time-dependent fields as well. 
It can be extended further in several directions: 
to allow for an arbitrary relative orientation of the particle velocity and magnetic moment; 
to account for non-Gaussianity of the displacement distribution; 
and to incorporate orientation-dependent interparticle interactions within the framework of the dynamic mean-field theory~\cite{Ilg2005,IvanovCamp2018}. 
The method could also potentially be employed to study other forms of active transport with aligning torques,\textit{ e.g.}, gravitaxis~\cite{tenHagen2014,Chepizhko2022} and chemotaxis~\cite{Taktikos2011,Lutier2026}.


\section{Methods} \label{sec:methods}

\paragraph{FPE numerical solution.}

\newcommand{\vctr}[1]{\bm{\mathrm{#1}}}

As detailed in the Supplementary Note 1, 
FPE~(\ref{eq:fpe}) produces the following vector ODEs:

\begin{subequations} \label{eq:fpe_vec}
\begin{equation} 
    \frac{d \vctr{c}^n}{d \tau} = \vctr{A}_0 \cdot \vctr{c}^n + {2\per} n \: \vctr{B}^{\parallel} \cdot \vctr{c}^{n-1} + \difc n (n-1) \: \vctr{c}^{n-2}, 
\end{equation}
\begin{multline}
    \frac{d \vctr{d}^n_m}{d \tau} = \vctr{A}_m \cdot \vctr{d}^n_m + \difc (n^2 - m^2) \vctr{d}^{n-2}_m + \\ {2\per} \left[ (n+m) \vctr{B}^{\perp}_{m}\cdot \vctr{d}^{n-1}_{m-1} - (n-m)\vctr{B}^{\perp}_{-m}\cdot \vctr{d}^{n-1}_{m+1}\right] ,
\end{multline}   
\end{subequations}

\noindent 
where $\vctr{c}^n = [c^n_0, c^n_1, \cdots, c^n_L]$ and 
$\vctr{d}^n_m = [0, \cdots, 0, d^n_{|m|,m}, \cdots, d^n_{L,m}]$ 
are vectors of size $L + 1$, 
$\vctr{A}_m$, $\vctr{B}^{\parallel}$ and $\vctr{B}^{\perp}_m$ 
are $(L+1) \times (L + 1)$ constant tridiagonal matrices,

\begin{gather}
\vctr{A}_m = \mathrm{tridiag}\Big( \mathfrak{a}^{-}_{l,m} ; \mathfrak{a}_{l} ; \mathfrak{a}^{+}_{l,m}\Big), \:\:\: l = 0, 1, \cdots, L, \nonumber \\ 
\vctr{B}^{\parallel} = \mathrm{tridiag}\Big( \mathfrak{b}^{\parallel,-}_{l}; 0;  \mathfrak{b}^{\parallel,+}_{l}\Big), \nonumber \\
\vctr{B}^{\perp}_m = \mathrm{tridiag}\Big( \mathfrak{b}^{\perp,-}_{l,m}; 0;  \mathfrak{b}^{\perp,+}_{l,m}\Big), \nonumber \\
\mathfrak{a}_{l} = -\frac{l(l+1)}{2}, \nonumber \\ 
\mathfrak{a}^{-}_{l,m} = \frac{\xi(l+1)\sqrt{l^2 - m^2}}{2(2l+1)}, \:\:
\mathfrak{a}^{+}_{l,m} = - \frac{\xi l\sqrt{(l+1)^2 - m^2}}{2(2l+1)},  \nonumber \\
\mathfrak{b}^{\parallel, -}_{l} = \frac{l}{2(2l+1)}, \:\: \mathfrak{b}^{\parallel, +}_{l} = \frac{l + 1}{2(2l+1)}, \nonumber \\
\mathfrak{b}^{\perp, -}_{l,m} = \frac{\sqrt{(l+m)(l+m-1)}}{4(2l+1)},\nonumber \\ 
\mathfrak{b}^{\perp, +}_{l,m} = -\frac{\sqrt{(l-m+1)(l-m+2)}}{4(2l+1)} \nonumber.
\end{gather}

\noindent 
Translational PDF moments are obtained as $\av{x^n_{\parallel}} = c^n_0$ and $\av{x_\perp^{2n}} = d^{2n}_{0,0}$. 
Initial conditions are ${c}^0_l(0) = {d}^0_{l,0}(0) = \la_l(\xi)$,
$\la_{l-1}(\xi) - \la_{l+1}(\xi) = (2l + 1)\la_l(\xi)/\xi$,
$\la_0(\xi) = 1$, $\la_1(\xi) = \la(\xi)$,
$c^n_l(0) = 0$ for $n \ge 1$ 
and $d^n_{l,m}(0) = 0$ for $n \ge 1$ or $m \neq 0$. 
The exact FPE solution is retrieved in the limit $L \rightarrow \infty$.
In practice, we found that $L = 30$ was sufficient to accurately obtain
the first four moments.
Eq.~(\ref{eq:fpe_vec}) was solved numerically using \texttt{solve\_ivp} function 
from \texttt{SciPy} Python library.

\paragraph{Effective field approximation.}
Consider the space-averaged PDF $Q(\ov,\tau) = \int W d \pv$. 
In thermodynamic equilibrium, it must reduce to the Boltzmann distribution Eq.~(\ref{eq:q_eq}).
The key idea behind EFA is to assume that in any non-equilibrium state $Q$ still has a Boltzmann-like shape, but instead of an actual field $\bm{\xi}$, 
it depends on an effective one~$\bm{\ef}$~\cite{martsenyuk1974kinetics,raikher1994effective}:
\begin{equation}
    Q_\text{EFA} =  \frac{\ef}{4\pi \sinh \ef} \exp({\bm{\ef}} \cdot \ov ). 
\end{equation}

\noindent 
The time-dependent effective field $\bm{\ef} = \bm{\ef}(\tau)$ is an unknown parameter to be determined self-consistently from the FPE. 
In our case, the external field imposes an {axial symmetry},  
so we can simplify the task by setting $\bm{\ef} = \ef \, \fv$.    

For transport properties, the method obviously has to be expanded to incorporate translational degrees of freedom. 
In the passive limit ($\per = 0$), rotational and translational motion are decoupled, and the orientation-averaged PDF $P(\pv,\tau) = \int W d \ov$
must take the standard Gaussian form with zero mean and constant diffusion coefficient.
It is also known that for ABPs without rotational bias ($\xi = 0$), MSD can be described in terms of an effective diffusivity taking different values at different time scales~\cite{bechinger2016active}. 
So, for magnetic ABPs we attempt the Gaussian ansatz
\begin{equation}
    P_\text{EFA}= \frac{1}{(4\pi \tau)^{3/2} \ed_{\parallel}^{1/2} \ed_{\perp}}\exp\left[ - \frac{(x_{\parallel} - \langle x_{\parallel}\rangle)^2}{4\ed_{\parallel}\tau} - \frac{x_{\perp}^2}{4 \ed_{\perp} \tau}\right].
\end{equation}

\noindent
Here, 
$\ed_{\parallel} = \ed_{\parallel}(\tau)$ and $\ed_{\perp} = \ed_{\perp}(\tau)$ are longitudinal and transverse \textit{ time-dependent} effective diffusivities, respectively.
We allow for a non-zero first moment $\langle x_{\parallel} \rangle$, while demanding zero mean displacements in the orthogonal plane.

Using the product $P_\text{EFA}\:Q_\text{EFA}$ as a PDF approximation is sufficient to derive Eq.~(\ref{eq:drift}) for the first moment.
However, it is not enough to account for the effect of activity on the second moments.
So, as a final ansatz we propose the function
\begin{equation} \label{eq:pdf_ef}
    W_\text{EFA} = P_\text{EFA} \:Q_\text{EFA} \Big[ 1 + \ecp_{\parallel} \delta {x}_{\parallel}\delta {n}_{\parallel}  + \ecp_{\perp} ( x_1 n_1 + x_2 n_2) \Big], 
\end{equation}

\noindent
where 
$\delta x_{\parallel} = x_{\parallel} - \langle x_{\parallel}\rangle$, 
$\delta n_{\parallel} = n_{3} - \langle n_{3}\rangle$,
$\ecp_{\perp} = \ecp_{\perp}(\tau)$ and $\ecp_{\parallel} = \ecp_{\parallel}(\tau)$ are the rotation-translation coupling parameters.
The coupling terms tie together rotational and translational fluctuations to account for an obvious tendency of ABPs to move in the direction they are facing.
These terms do not change the shape of $P$ or $Q$ and do not affect the PDF normalization.
However, they ensure that $\langle x_i  n_j n_k \ldots \rangle \neq \langle x_i \rangle \langle n_j n_k \ldots \rangle$, which turns out to be crucial to capture the effects of activity on diffusion.
As shown in the Supplementary Note 2, 
substitution of Eq.~(\ref{eq:pdf_ef}) into FPE~(\ref{eq:fpe}) 
gives in an unbounded domain

\begin{equation}
     \ed_{\rho}(\tau) = \difc_{\rho}^\mathrm{eff} - \Big(\difc_{\rho}^\mathrm{eff} - \ed_\rho(0)\Big) \left( 1 - e^{-\tau/\tau_{\rho}} \right)\frac{\tau_{\rho}}{\tau},
\end{equation}

\noindent
where $\rho = \:\parallel$ or $\perp$, $\difc_{\rho}^\mathrm{eff} = \Delta \difc_{\rho}  + \difc$, $\tau_{\rho}$ and $\Delta \difc_{\rho}$ are given by Eqs.~(\ref{eq:taus}) and (\ref{eq:delta_ds}), respectively. 
Eq.~(\ref{eq:msd}) is then obtained from known properties of the Gaussian distribution, assuming the initial condition $\ed_\rho(0) = \difc$.
For the coupling parameter one obtains

\begin{equation}
    \ecp_{\rho}(\tau) =  \frac{\per}{2 \ed_{\rho}(\tau)} \left(1 - e^{-\tau/\tau_{\rho}}\right) \frac{\tau_{\rho}}{\tau},
\end{equation}

Note, that our $\Delta \mathcal{D}_{\rho}$  is a direct analog of Eq. (37) in Ref.~\cite{VidalUrquiza2017}. 
The difference comes down to specific expressions used for relaxation times $\tau_{\rho}$.
We reobtain EFA expressions, first found in Ref.~\cite{martsenyuk1974kinetics}, 
while  Ref.~\cite{VidalUrquiza2017} used exact inifinite series representations, first obtained in Ref.~\cite{Waldron1994}. 
Exact and EFA diffusivities have the same asymptotics at ${\xi \gg 1}$, ${\Delta \mathcal{D}_{\parallel}(\xi)/ \Delta \mathcal{D}_{\parallel}(0) = 3/\xi^3}$, ${\Delta \mathcal{D}_{\perp}(\xi)/\Delta \mathcal{D}_{\perp}(0) = 6/\xi^2}$. 
Weak-field expansions, ${\Delta \mathcal{D}_{\rho}(\xi)/ \Delta \mathcal{D}_{\rho}(0) = 1 - r_{\rho}\xi^2}$, 
are different. 
Exact coefficients are $r_{\parallel} = 14/45$ and $r_{\perp} = 3/20$~\cite{Takatori2014,VidalUrquiza2017}, 
while the EFA gives $r_{\parallel} = 1/3$ and $r_{\perp} = 1/6$.

\paragraph{Brownian dynamics.}

Brownian dynamics simulations were performed using the software package \texttt{ESPResSo~4.2}~\cite{espresso4}.
Eqs.~(\ref{eq:lang_nd}) were numerically integrated using the standard Brownian thermostat of \texttt{ESPResSo} with the dimensionless time step $\Delta \tau = 10^{-3}$.

\paragraph{First-passage problem.}

For a Gaussian distribution with constant diffusion coefficients $\difc_{\parallel, \perp}^\mathrm{eff}$ and a constant drift velocity $u_{\parallel} \ge 0$, the first-passage problem in a semi-infinite domain can be straightforwardly solved using the standard method of images~\cite{Redner2001,MOLINI2011}. 
Let's say, that the vertical coordinate of the target is $x_{\parallel} = 0$.
Then the displacement distribution inside the domain is

\begin{multline}
    P_\text{FPP}(x_{\parallel} \le 0, x_{\perp}, \tau) = \frac{1}{(4\pi \tau)^{3/2}} \frac{1}{\difc_{\perp}^\mathrm{eff} \sqrt{\difc_{\parallel}^\mathrm{eff}}} e^{-x_{\perp}^2/4 \difc_{\perp}^\mathrm{eff} \tau} \\
    \times \left[ 
    e^{-(x_{\parallel} + \ell_{\parallel} - u_{\parallel}\tau)^2/4 \difc_{\parallel}^\mathrm{eff} \tau} - w \, e^{-(x_{\parallel} - \ell_{\parallel} - u_{\parallel}\tau)^2/4 \difc_{\parallel}^\mathrm{eff} \tau} \right],
\end{multline}

\noindent
where $w = \exp \Big(\ell_{\parallel} u_{\parallel}/\difc^\text{eff}_{\parallel}\Big)$ is the weight of the image Gaussian, which ensures that the absorbing boundary condition is always fulfilled, $P_\text{FPP}(0, x_{\perp}, \tau) = 0$.
The survival probability, Eq.~(\ref{eq:surv}), 
then follows from 

\begin{equation}
    \mathcal{S}(\tau) = 2\pi \int_0^\infty \int_{-\infty}^0 x_\perp  P_\text{FPP} \, d x_{\parallel} d x_{\perp}. 
\end{equation}

\noindent
The first passage probability is $f_\text{FPP} = - d \mathcal{S}/d \tau$ and
the mean inverse absorption time, Eq.~(\ref{eq:inv_time}), is calculated from

\begin{equation}
    \av{\tau^{-1}_\text{abs}} = \int_0^{\infty}\frac{f_\text{FPP}}{\tau}  d \tau.
\end{equation}

\noindent
The normal component of the particle flux at the target surface is

\begin{equation}
    j_\text{FPP} ( \ell_{\perp}, \tau) = u_{\parallel} P_\text{FPP}\Bigg|_{\genfrac{}{}{0pt}{}{x_{\parallel} = 0\hfill}{x_\perp = \ell_{\perp}}\hfill} - \difc_{\parallel}^\mathrm{eff} \frac{\partial P_\text{FPP}}{\partial x_{\parallel}}\Bigg|_{\genfrac{}{}{0pt}{}{x_{\parallel} = 0\hfill}{x_\perp = \ell_{\perp}\hfill}},
\end{equation}

\noindent
and the long-term surface distribution of particles is 

\begin{equation}
    \mathcal{E}(\ell_{\perp}) = \int_0^\infty j_\text{FPP} \, d \tau.    
\end{equation}

\noindent
The mean sine of the spread angle, Eq.~(\ref{eq:sin}), 
is finally obtained from

\begin{equation}
    \av{\sin \psi} = 2\pi \int_0^{\infty} \frac{\ell^2_{\perp}}{\sqrt{\ell_{\perp}^2 + \lfp^2}} \, \mathcal{E} d \ell_{\perp}.
\end{equation}

\section*{Acknowledgments}
This research was funded in part by the Austrian Science Fund (FWF) [10.55776/PAT4307624].
For open access purposes, the authors have applied a CC BY public copyright license to any author-accepted manuscript version arising from this submission.
V.S. acknowledges support from the European Union's Horizon Europe research and innovation programme under the Marie Sk{\l}odowska-Curie Actions (MSCA), grant agreement number 101149450.
A.V.C. acknowledges support from the BMFTR project 01DK24006 PLASMA-SPIN-ENERGY.

\section*{Author contributions}
{A.A.K. and V.S. designed the research based on the original suggestion from S.S.K. 
A.A.K. developed the methodology and performed all derivations and calculations.
V.S., S.S.K. and A.V.C. contributed to the discussion and interpretation of the results.
A.A.K. and V.S. wrote the first draft.
All authors reviewed and edited the final version of the manuscript.}


\bibliography{references}

\clearpage

\onecolumngrid 

\setcounter{equation}{0}

\renewcommand{\thesection}{S\arabic{section}}
\renewcommand{\theequation}{S\arabic{equation}}
\renewcommand{\thefigure}{S\arabic{figure}}
\renewcommand{\thetable}{S\arabic{table}}

\renewcommand{\theHequation}{supp.equation.\arabic{equation}}

\begin{center}
{\large\bfseries Supplemental Material for}\\[1em]
{\bfseries ``Drift-diffusion interplay in active Brownian particles under orienting field''}
\end{center}



\section*{Supplementary Note 1: Direct numerical solution of the Fokker-Planck equation} \label{sec:fpe}

\subsection{Probability density function and its spatial moments}

While the full probability distribution function $W = W(\pv,\ov,\tau)$ depends on both particle position and orientation, of a more practical interest is the rotation-averaged PDF, 
that gives the probability of finding ABP 
at a point $\bm{x}$ regardless of its orientation,

\begin{equation}
    P(\bm{x},\tau) = \int W(\bm{x}, \hat{\bm{n}}, \tau) d^2 \hat{\bm{n}}, \:\:\: P(\bm{x},0) = \delta(\bm{x}).
\end{equation}

\noindent Going forward, we can make use of the cylindrical symmetry introduced by the field. 
If we choose the Cartesian reference frame in which $\hat{\bm{h}} = [0,0,1]$, 
PDF should only depend on the longitudinal distance along the field
and on the radial distance in the orthogonal plane, 
\textit{i.e.}, $P(\bm{x}, \tau) = P(x_{\parallel},x_{\perp},\tau)$,
where
\begin{equation}
x_{\parallel} = x_3, \:\:\: x_{\perp} = \sqrt{x_1^2 + x_2^2}.
\end{equation}
One of the main points of interest in this work is the temporal evolution 
of PDF longitudinal and orthogonal spatial moments,

\begin{equation}
\left\langle x^n_{\parallel} \right\rangle(\tau)  = \int x^n_{\parallel} P(\bm{x}, \tau) d^3 \bm{x}, \:\:\: 
\left\langle x^{2n}_{\perp} \right\rangle(\tau) = \int x^{2n}_{\perp} P(\bm{x}, \tau) d^3 \bm{x}, \:\:\: 
n \ge 1. \label{eq:mom_def}
\end{equation}

\noindent Note that due to aforementioned cylindrical symmetry, odd orthogonal Cartesian moments must vanish, $\langle x^{2n - 1}_{1}\rangle = \langle x^{2n - 1}_{2}\rangle = 0$. So, we will only focus on even moments.

A useful object that will help us with obtaining the moments is the intermediate scattering function (ISF) -- that is, 
the Fourier transform of PDF:

\begin{align}
    F(\bm{k}, \tau) &= \int e^{-i \bm{k} \cdot \bm{x}} P(\bm{x},\tau) d^3 \bm{x} \notag \\
    &= \int^{\infty}_{-\infty} \int^{\infty}_{0} \int^{2\pi}_{0} 
    e^{-i {k}_{\parallel} {x}_{\parallel}} e^{-i {k}_{\perp} {x}_{\perp} \cos(\phi_x - \phi_k )} P(x_{\parallel},x_{\perp},\tau) x_{\perp} d x_{\parallel} dx_{\perp} d\phi_x \notag \\
    &= 2\pi \int^{\infty}_{-\infty} \int^{\infty}_{0} 
    e^{-i {k}_{\parallel} {x}_{\parallel}} J_0({k}_{\perp} {x}_{\perp} ) P(x_{\parallel},x_{\perp},\tau) x_{\perp} d x_{\parallel} dx_{\perp} \notag \\
    &= F(k_{\parallel}, k_{\perp}, \tau),  \label{eq:isf}
\end{align}

\noindent where $\bm{k} = \left[k_{\perp} \cos (\phi_k),k_{\perp}\sin(\phi_k),k_{\parallel}\right]$ is the Fourier wave vector, 
$J_0(x) = \int^{2\pi}_0 e^{- i x \cos \phi} d\phi/2\pi$ is the Bessel function of the first kind of order zero. 
From (\ref{eq:mom_def}) and (\ref{eq:isf}), 

\begin{gather}
    \left\langle x^{n}_{\parallel} \right\rangle (\tau) = i^n \frac{\partial^n F(k_{\parallel},k_{\perp},\tau)}{\partial k_{\parallel}^n } \Bigg|_{\bm{k} = \bm{0}}, \\
    \left\langle x^{2n}_{\perp} \right\rangle (\tau) = (-1)^n 2^{2n}\frac{(n!)^2}{(2n)!}\frac{\partial^{2n} F(k_{\parallel},k_{\perp},\tau)}{\partial k_{\perp}^{2n} } \Bigg|_{\bm{k} = \bm{0}}. 
\end{gather}

\subsection{Fourier transform of the Fokker-Planck equation}




Upon the spatial Fourier transform, Fokker-Planck equation [Eq. (4) in the main text] takes the form

\begin{equation} \label{eq:fpe_fourier}
2 \frac{\partial \widetilde{W}}{\partial \tau}  = - 2 i {\per} \left(\hat{\bm{n}} \cdot \bm{k} \right) \widetilde{W} - {2\difc} \bm{k}^2 \widetilde{W} - \xi \bm{\mathcal{R}}_{\ov} \cdot \left( [\hat{\bm{n}} \times \hat{\bm{h}}]\widetilde{W}\right) + \bm{\mathcal{R}}^2_{\ov} \widetilde{W},
\end{equation}

\noindent where

\begin{equation}
    \widetilde{W}(\bm{k},\hat{\bm{n}},\tau) = \int {W}(\bm{x},\hat{\bm{n}},\tau) e^{- i \bm{x} \cdot \bm{k}} d^3 \bm{x}.
\end{equation}

\noindent Next, we rewrite FPE in terms of ABP orientation angles ($\hat{\bm{n}} = [\sin \theta \cos \phi,\sin \theta \sin \phi , \cos \theta]$ and $y = \cos\theta$):

\begin{align} \label{eq:fpe_angles}
2 \frac{\partial \widetilde{W}}{\partial \tau}  = &- 2 i  {\per} \left(\sqrt{1 - y^2} k_{\perp}\cos(\phi - \phi_k) + y k_{\parallel}\right) \widetilde{W} - {2\difc} \left({k}^2_{\perp} + {k}^2_{\parallel}\right)\widetilde{W} \notag \\
&+ \frac{\partial}{\partial y} \Bigg[ (1-y^2) \left( \frac{\partial\widetilde{W}}{\partial y} - \xi \widetilde{W}\right)\Bigg] + \frac{1}{1 - y^2}\frac{\partial^2 \widetilde{W}}{\partial \phi^2}.
\end{align}

\noindent To get rid of explicit angular dependencies, 
$\widetilde{W}$ can be expanded into a series of normalized spherical harmonics

\begin{equation} \label{eq:sh_exp}
    \widetilde{W}(\bm{k},\hat{\bm{n}},\tau) = \sum_{l = 0}^{\infty} \sum_{m = -l}^{m=l} b_{l,m}(\bm{k}, \tau) Y_{l,m}(y,\phi),
\end{equation} 

\noindent which are defined here as

\begin{equation} 
    Y_{l,m}(y,\phi) = (-1)^{m} \sqrt{\frac{(2l+1)}{4\pi}\frac{(l-m)!}{(l+m)!}}P^m_l(y)e^{i m\phi}
\end{equation}

\noindent with $P^m_l$ being the associated Legendre polynomials and $-l \le m \le l$.
Harmonics have properties

\begin{equation} \label{eq:sh_prop}
 Y_{l,m}^* = (-1)^mY_{l,-m}, \:\:\: \int Y_{l,m} Y^*_{l',m'} d^2\hat{\bm{n}} = \delta_{ll'}\delta_{mm'},
\end{equation}

\noindent and the expansion coefficients in Eq.~(\ref{eq:sh_exp})
can be formally found as 

\begin{equation}
    b_{l,m}(\bm{k},\tau) = \int Y^*_{l,m} \widetilde{W}(\bm{k},\hat{\bm{n}},\tau) d^2 \hat{\bm{n}}.
\end{equation}

\noindent It is easy to check that $F(\bm{k}, \tau) = \sqrt{4 \pi} \: b_{0,0}(\bm{k}, \tau)$.
Upon inserting expansion Eq.~(\ref{eq:sh_exp}) into Eq.~(\ref{eq:fpe_angles}) and averaging over orientations with the account of  orthogonality property Eq.~(\ref{eq:sh_prop}), 
we obtain the following infinite set of differential recurrence relations
for PDF Fourier coefficients:

\begin{align}
2 \frac{\partial b_{l,m}}{\partial \tau} + \left[l(l + 1) + {2\difc} (k_{\perp}^2 + k_{\parallel}^2)\right] b_{l,m} & = &  &\notag\\
- i{\per} k_{\perp} e^{-i \phi_k} \Bigg[& \sqrt{\frac{(l + m - 1)(l + m)}{(2l - 1)(2l+1)}} b_{l-1,m-1} &-\:\:\:   \sqrt{\frac{(l - m + 1)(l - m + 2)}{(2l + 1)(2l+3)}}  b_{l+1,m - 1} & \Bigg]\notag \\
+ i{\per} k_{\perp} {e^{i \phi_k}}  \Bigg[& \sqrt{\frac{(l - m -1)(l-m)}{(2l - 1)(2l+1)}} b_{l-1,m+1} &-\:\:\:\sqrt{\frac{(l + m + 1)(l + m + 2)}{(2l + 1)(2l+3)}}  b_{l+1,m + 1} & \Bigg]\notag \\
- 2i{\per} k_{\parallel}  \Bigg[& \sqrt{\frac{(l - m)(l+m)}{(2l - 1)(2l+1)}} b_{l-1,m} &+\:\:\:  \sqrt{\frac{(l - m  + 1)(l+ m + 1)}{(2l + 1)(2l+3)}}  b_{l+1,m} & \Bigg]\notag \\
+  \xi \Bigg[& (l+1)\sqrt{\frac{(l - m)(l+m)}{(2l - 1)(2l+1)}}  b_{l-1,m} &-\:\:\:l \sqrt{\frac{(l  - m + 1)(l+ m + 1)}{(2l + 1)(2l+3)}}  b_{l+1,m} &\Bigg]. \label{eq:fpe_coefs}
\end{align}

\noindent Note that for passive particles (${\per} = 0$) in zero field ($\xi = 0$),
equations for different coefficients are independent, 
while both field and self-propulsion introduce coupling between the nearest-neighbor harmonics.

\subsection{Moment equations} \label{sec:mom_eqs}

If one is interested in spatial moments, Eq.~(\ref{eq:fpe_coefs}) can be further simplified.
For instance, to consider longitudinal moments $\langle x^n_{\parallel}\rangle$ we can first set $k_{\perp}$ to zero --
it will remove the coupling between different $m$ values and we can focus only on $m = 0$.
Then differentiating Eq.~(\ref{eq:fpe_coefs}) $n$ times over $k_{\parallel}$ and taking $k_{\parallel} = 0$ produces a new differential recurrence set:


\begin{equation}
    2 \frac{d c^n_{l}}{d \tau} +  l(l+1)c^n_{l} = 
    \xi \frac{l(l+1)}{2l + 1} \left( c^n_{l-1}  - c^n_{l+1} \right) + {2\per} n \left( \frac{l }{{2l+1} }c^{n-1}_{l-1}  + \frac{l+1}{{2l+1}}c^{n-1}_{l+1}\right) + {2\difc} n(n - 1) c^{n-2}_l , \label{eq:mom_par}
\end{equation}

\noindent where 

\begin{equation}
c^n_l(\tau) = \sqrt{\frac{4\pi} {2l + 1}} i^n \Bigg[\frac{\partial^n }{\partial k_{\parallel}^n} b_{l,0}(\bm{k},\tau)\Bigg]_{\bm{k} = \bm{0}}
\end{equation}

\noindent for $n \ge 0$, $l \ge 0$ and $c^n_l = 0$ otherwise. 
Evidently, $c^n_0 = \langle x_{\parallel}^n\rangle$.
More broadly, $c^n_l = \left\langle x_{\parallel}^n P_l(n_{\parallel})\right\rangle$,
where $P_l(n_{\parallel} ) = P^0_l(n_{\parallel} )$ are ordinary Legendre polynomials,
$n_{\parallel} = \hat{\bm{n}}\cdot\hat{\bm{h}} = y$ 
is the projection of the orientation vector on the field direction.
Since $P_1(n_{\parallel} ) = n_{\parallel}$, 
coefficient $c^0_1 = \langle n_{\parallel} \rangle$ 
represents the space-averaged dipole moment of the ABP ensemble.
This quantity -- and in fact all the space-averaged orientation harmonics $c^0_l$ --
is not affected by the particle self-propulsion as can be seen from Eq.~(\ref{eq:mom_par}) upon setting $n = 0$.
A well known result from the statistical physics of dilute ferrofluids
is that in a static field moments $c^0_l$ will always approach equilibrium values,
which can be written in terms of the so-called \textit{Langevin~function}~$\mathcal{L}(\xi)  = \coth \xi - 1/\xi$,

\begin{equation} \label{eq:c0l}
    c^0_l(\tau \gg 1) = \frac{\int^1_{-1}P_l(y)e^{\xi y} dy}{\int^1_{-1}e^{\xi y} dy} = \frac{I_{l+1/2}(\xi)}{I_{1/2}(\xi)} = \mathcal{L}_l(\xi).
\end{equation}

\noindent Here 
$I_l(\xi)$ are the modified Bessel functions of the first kind of order~$l$,
$\mathcal{L}_0 = 1$, $\mathcal{L}_1(\xi) = \mathcal{L}(\xi)$,
functions $\mathcal{L}_{l>1}(\xi)$ can be found from the recurrence relation
$\mathcal{L}_{l-1}(\xi) - \mathcal{L}_{l+1}(\xi) = (2l + 1)\mathcal{L}_l(\xi)/\xi$ [it follows from Eq.~(\ref{eq:mom_par}) upon $n = 0$ and $d/d\tau = 0$].
It is easy to check that our choice of the initial conditions [see the ``Model'' section of the paper] 
means that $c^0_l(\tau = 0) = \mathcal{L}_l(\xi)$.
In~other words, space-averaged orientation harmonics are set to their equilibrium values from the beginning and therefore must remain constant throughout the ABP motion.
On the contrary, moments with $n \ge 1$ can evolve in time starting from zero initial conditions, $c^{n\ge 1}_l(0) = 0$.

To get the orthogonal moment equation, 
Eq.~(\ref{eq:fpe_coefs}) can be first multiplied by $e^{im\phi_k}$
and then averaged over $\phi_k$. 
Differentiating the result $n$ times over $k_{\perp}$ and taking $\bm{k} = \bm{0}$
produces

\begin{align}
2 \frac{\partial d^n_{l,m}}{\partial \tau} + l(l + 1)d^n_{l,m}  =  {2\difc} & (n^2 - m^2)  d^{n-2}_{l,m} &  &\notag\\
+   \frac{\xi}{2l + 1} \Bigg[& (l+1) \sqrt{(l-m)(l+m)}d^n_{l-1,m}  &- \:\:\: l \sqrt{(l-m+1)(l+m+1)} d^n_{l+1,m} &\Bigg] \notag\\
+ \frac{{\per} }{(2l + 1)} \Bigg[& (n+m)\sqrt{(l+m)(l+m-1)} d^{n-1}_{l-1,m-1} &- \:\:\:(n+m) \sqrt{(l - m + 1)(l - m + 2)}  d^{n-1}_{l+1,m - 1} & \notag \\
 - &(n-m)\sqrt{(l - m -1)(l-m)} d^{n-1}_{l-1,m+1}  &+ \:\:\: (n-m)\sqrt{(l + m + 1)(l + m + 2)} d^{n-1}_{l+1,m + 1}  &\Bigg],
 \label{eq:mom_per}
\end{align}

\noindent where coefficients $d^n_{l,m}$ are different from zero for $n \ge 0$, $l \ge 0$, $|m| \le \text{min}(l,n)$ and for $n$ and $m$ having the same parity. In this case they are given by  
\begin{align}
    d^n_{l,m}(\tau) &= 2^n (-1)^{\frac{m-n}{2}} e^{i \pi m/2} 
    \binom{n}{\frac{n - m}{2}}^{-1}\sqrt{\frac{4 \pi}{2 l + 1}}
    \Bigg[\frac{\partial^n  }{\partial k_{\perp}^n} \langle b_{l,m}(\bm{k}, \tau)e^{im\phi_k} \rangle_{\phi_k} \Bigg]_{\bm{k} =\bm{0}} \notag \\
    &=  (-1)^m \sqrt{\frac{(l-m)!}{(l+m)!}} \left\langle P^m_{l}e^{im(\phi_x - \phi)}x^n_{\perp}\right\rangle(\tau),
\end{align}

\noindent where

\begin{align}
    \langle b_{l,m} e^{im\phi_k}  \rangle_{\phi_k} &= \frac{1}{2\pi} \int^{2\pi}_0 b_{l,m}(\bm{k}, \tau) e^{im\phi_k} d \phi_k \notag \\
    &= \int\int Y^*_{l,m} e^{-i k_{\parallel} x_{\parallel}} e^{im\phi_x}e^{-i\pi m/2}J_{m}(x_{\perp} k_{\perp}) W d^2\hat{\bm{n}}  d^3 \bm{x}.
\end{align}

\noindent We also used the fact that for $-n \le m \le n$
\begin{equation}
    \frac{\partial^n J_m(x_{\perp} k_{\perp})}{\partial k_{\perp}^n } \Bigg|_{k_{\perp} = 0} =
    \begin{cases}
    2^{-n} (-1)^{\frac{n-m}{2}} \binom{n}{\frac{n - m}{2}}x^n_{\perp},& \text{if $n$ and $m$ have the same parity,}\\
    0,              & \text{otherwise.}
\end{cases}
\end{equation}
\noindent One can see that $d^{2n}_{0,0} = \langle x^{2n}_{\perp}\rangle$.
Coefficients satisfy the initial conditions $d^0_{l,0} (\tau = 0) = \mathcal{L}_l(\xi)$ and $d^{n}_{l, m}(0) = 0$ for $n \ge 1$ or $m \neq 0$.
They also have the property $ (d^n_{l,m})^* = (-1)^m d^n_{l,-m}$.

\subsection{Solving the moment equations}

While the number of harmonics in the expansion Eq.~(\ref{eq:sh_exp}) is infinite, 
for numerical evaluation we will inevitably have to restrict ourselves 
with some large but finite set. 
To take that into account, we introduce the cutoff value $L$ and set $c^n_{l > L} = 0$, $d^n_{l> L, m} = 0$. 
An exact solution is obtained in the limit $L \rightarrow \infty$.

Eqs.~(\ref{eq:mom_par}) and (\ref{eq:mom_per}) can be rewritten in vector forms:

\begin{gather} \label{eq:vec_long}
    \frac{d \vctr{c}^n}{d \tau} = \vctr{A}_0 \cdot \vctr{c}^n + {2\per} n \: \vctr{B}^{\parallel} \cdot \vctr{c}^{n-1} + \difc n (n-1) \: \vctr{c}^{n-2}, \\
    \frac{d \vctr{d}^n_m}{d \tau} = \vctr{A}_m \cdot \vctr{d}^n_m + {2\per} \left[ (n+m) \vctr{B}^{\perp}_{m}\cdot \vctr{d}^{n-1}_{m-1} - (n-m)\vctr{B}^{\perp}_{-m}\cdot \vctr{d}^{n-1}_{m+1}\right] + \difc (n^2 - m^2) \vctr{d}^{n-2}_m,
\end{gather}

\noindent where $\vctr{c}^n = [c^n_0, c^n_1, \cdots, c^n_L]$ and 
$\vctr{d}^n_m = [0, \cdots, 0, d^n_{|m|,m}, \cdots, d^n_{L,m}]$ 
are vectors of size $L + 1$, 
$\vctr{A}_m$, $\vctr{B}^{\parallel}$ and $\vctr{B}^{\perp}_m$ 
are $(L+1) \times (L + 1)$ constant tridiagonal matrices,

\begin{gather}
\vctr{A}_m = \begin{bmatrix}
\mathfrak{a}_{0} & \mathfrak{a}^{+}_{0,m} & \cdots & 0 & 0  \\
\mathfrak{a}^{-}_{1,m} & \mathfrak{a}_{1} &  \cdots & 0 & 0  \\
\vdots & \vdots &  \ddots & \vdots & \vdots \\
0 & 0 &  \cdots  & \mathfrak{a}_{L-1} & \mathfrak{a}^{+}_{L-1,m} \\
0 & 0 & \cdots  & \mathfrak{a}^{-}_{L,m} & \mathfrak{a}_{L}
\end{bmatrix}, \:
\vctr{B}^{\parallel} = \begin{bmatrix}
0 & \mathfrak{b}^{\parallel,+}_{0} & \cdots & 0 & 0  \\
\mathfrak{b}^{\parallel,-}_{1} & 0 &  \cdots & 0 & 0  \\
\vdots & \vdots &  \ddots & \vdots & \vdots \\
0 & 0 &  \cdots  & 0 & \mathfrak{b}^{\parallel,+}_{L-1} \\
0 & 0 & \cdots  & \mathfrak{b}^{\parallel, -}_{L} & 0
\end{bmatrix}, \:
\vctr{B}^{\perp}_m = \begin{bmatrix}
0 & \mathfrak{b}^{\perp,+}_{0,m} & \cdots & 0 & 0  \\
\mathfrak{b}^{\perp,-}_{1,m} & 0 &  \cdots & 0 & 0  \\
\vdots & \vdots &  \ddots & \vdots & \vdots \\
0 & 0 &  \cdots  & 0 & \mathfrak{b}^{\perp,+}_{L-1,m} \\
0 & 0 & \cdots  & \mathfrak{b}^{\perp, -}_{L,m} & 0
\end{bmatrix}, \\
\mathfrak{a}_{l} = -\frac{l(l+1)}{2}, \:\: \mathfrak{a}^{-}_{l,m} = \frac{\xi(l+1)\sqrt{l^2 - m^2}}{2(2l+1)}, \:\:
\mathfrak{a}^{+}_{l,m} = - \frac{\xi l\sqrt{(l+1)^2 - m^2}}{2(2l+1)}, \nonumber \\
\mathfrak{b}^{\parallel, -}_{l} = \frac{l}{2(2l+1)}, \:\: \mathfrak{b}^{\parallel, +}_{l} = \frac{l + 1}{2(2l+1)},  \\
\mathfrak{b}^{\perp, -}_{l,m} = \frac{\sqrt{(l+m)(l+m-1)}}{4(2l+1)}, \:\:
\mathfrak{b}^{\perp, +}_{l,m} = -\frac{\sqrt{(l-m+1)(l-m+2)}}{4(2l+1)}. \nonumber
\end{gather}

\section*{Supplementary Note 2: Effective field approximation}

 \subsection{Equations for marginal distributions in an unbounded domain}

Equations for marginal distributions $Q(\ov, \tau) = \int W(\pv, \ov, \tau) d \pv$ and $P(\pv,\tau) = \int W(\pv,\ov,\tau) d \ov$ can be obtained by averaging the full Fokker-Planck equation over the particle position $\pv$ or orientation $\ov$, respectively, and then making use of the divergence theorem with the natural assumption that $W \rightarrow 0$ at $|\pv| \rightarrow \infty$. 
The resulting expressions are

\begin{gather}
    \frac{\partial Q}{\partial \tau} = - \frac{\xi}{2} \rop \cdot \left([\ov \times \fv] Q \right) + \frac{1}{2}\rop^2 Q, \label{eq:ofpe}\\
    \frac{\partial P}{\partial \tau} = -\per \bm{\nabla} \cdot \dir + \difc \bm{\nabla}^2 P. \label{eq:tfpe}
\end{gather}

\noindent
Equation~(\ref{eq:ofpe}) is a standard FPE describing magnetodynamics of a dilute ferrofluid with magnetically-rigid particles. 
It is not affected by translational degrees of freedom and activity.
On the contrary, in translational equation~(\ref{eq:tfpe}) activity introduces an explicit dependency on the average orientation vector per unit volume $\bm{\mathcal{N}} = \dir(\pv, \tau)$,

\begin{equation} \label{eq:dir}
    \dir = \int \ov W (\pv,\ov, \tau) d \ov.
\end{equation}

\noindent
Note that the vector field $\bm{\mathcal{M}}(\pv, \tau) = \mu \: \bm{\mathcal{N}}$ can be interpreted as an inhomogeneous ensemble magnetization.

\subsection{Mean orientation and mean displacement} \label{sec:efa_md}

From Eqs.~(\ref{eq:ofpe}) and (\ref{eq:tfpe}) one can obtain equations for the mean orientation and mean displacement of the particle, respectively:

\begin{gather}
    \frac{d}{d \tau} \langle n_i \rangle = \frac{\xi}{2} \Big( h_i - \langle n_i n_j\rangle h_j \Big) - \langle n_i \rangle, \label{eq:ode_n}\\
    \frac{d}{d \tau} \langle x_i \rangle = \per \langle n_i \rangle \label{eq:ode_x}.
\end{gather}

\noindent
Here, $i,j = 1, 2, 3$ and the summation over repeated indices is assumed. 
Eq.~(\ref{eq:ode_n}) contains $\langle n_i n_j\rangle$. 
Equation for the latter in turn will contain even higher order moments giving rise to an infinite chain of equations similar to what we saw in  Sec.~\ref{sec:fpe}.
The effective field ansatz, however, allows one to close this chain early. 
Indeed, using $Q_\text{EFA}$ and making use of the problem axial symmetry, we can write down

\begin{gather}
    \langle n_i \rangle = \la(\ef) h_i, \label{eq:efa_n} \\
    \langle n_i n_j \rangle = \frac{\la(\ef)} {\ef} \delta_{ij} + \la_2(\ef) h_i h_j, \label{eq:efa_nn} \\ 
    \langle n_i n_j n_k \rangle = \frac{\la_2(\ef)} {\ef} \Big(  h_i \delta_{jk} + h_j \delta_{ik} + h_k \delta_{ij}\Big)+ \la_3(\ef) h_i h_j h_k, \label{eq:efa_nnn}
\end{gather}

\noindent
where $\la_2(\ef) = 1 - 3\la(\ef)/\ef$ and $\la_3(\ef) = \la(\ef) - 5\la_2(\ef)/\ef$ [see Eq. (\ref{eq:c0l})].
Substituting Eqs.~(\ref{eq:efa_n}) and (\ref{eq:efa_nn}) into Eq.~(\ref{eq:ode_n}) gives

\begin{equation} \label{eq:ode_ef}
    \frac{d \ef}{d \tau} = \frac{\la(\ef)}{\la'(\ef)} \left(\frac{\xi}{\ef} - 1 \right).
\end{equation}

\noindent
In principle, Eqs.~(\ref{eq:ode_x}), (\ref{eq:efa_n}) and (\ref{eq:ode_ef}) allow one to calculate mean displacement and mean orientation for any time-dependent uniaxial field.
For a static field and equilibrium initial conditions, the solution is simply 

\begin{equation} \label{eq:efa_md}
    \ef = \xi, \:\:\: \langle \ov \rangle = \la(\xi) \fv, \:\:\: \langle \pv \rangle = \per \la(\xi)  \tau \fv.
\end{equation}

\subsection{Diffusion coefficients in a static field}

From Eq.~(\ref{eq:tfpe}), the equation that governs the second spatial moment is 

\begin{equation} \label{eq:ode_xx}
    \frac{d}{d \tau} \av{x_i x_j} = \per \Big( \av{x_i n_j} + \av{x_j n_i} \Big) + 2 \difc \delta_{ij}.
\end{equation}

\noindent It contains a \textit{joint moment} $\av{x_i n_j}$. 
Equation for it can only be obtained from the full FPE:

\begin{equation} \label{eq:ode_xn}
    \frac{d}{d \tau} \av{x_i n_j} = \per \av{n_i n_j}  + \frac{\xi}{2} \Big( \av{x_i}h_j - \av{x_i n_j n_k} h_k \Big) - \av{x_i n_j}.
\end{equation}

\noindent
Within the chosen effective field ansatz [Eq. (18) in the main text], MSD is given by 

\begin{equation} \label{eq:efa_xx}
    \langle x_i x_j \rangle = 2 \tau \ed_{ij} + \langle x_i\rangle \langle x_j\rangle,
\end{equation}

\noindent
and the joint moments are

\begin{equation} \label{eq:efa_xn}
    \langle x_i n_j n_k \ldots  \rangle = \langle x_i \rangle \langle n_j n_k \ldots \rangle + 2 \tau \ed_{il} \ecp_{lm} \left( \langle n_j n_k \ldots n_m\rangle -  \langle n_j n_k \ldots\rangle \langle n_m\rangle\right).
\end{equation}

\noindent
Here, we introduced diagonal matrices $\ed_{ij}$ and $\ecp_{ij}$ as

\begin{equation}
    \ed_{ij} = \begin{cases}
    \ed_{\parallel}, & \text{$i = j = 3$},\\
    0, & \text{$i \neq j$}, \\
    \ed_{\perp}, & \text{otherwise},
  \end{cases} 
  \:\:\:\:\:
    \ecp_{ij} = \begin{cases}
    \ecp_{\parallel}, & \text{$i = j = 3$},\\
    0, & \text{$i \neq j$}, \\
    \ecp_{\perp}, & \text{otherwise}.
  \end{cases} 
\end{equation}

\noindent
Substituting Eqs. (\ref{eq:efa_xx}), (\ref{eq:efa_xn}) into Eq. (\ref{eq:ode_xx}) and using results for orientational moments from Sec.~\ref{sec:efa_md}, one obtains equations for effective diffusion coefficients:

\begin{equation}
    \frac{d \ed_{\rho}}{d \tau} = 2 \per \Lambda_{\rho} \ecp_{\rho} \ed_{\rho} + \frac{\difc - \ed_{\rho}}{\tau}, \label{eq:ode_difc}
\end{equation}

\noindent
where $\rho = \: \parallel$ or $\perp$, 

\begin{equation}
\Lambda_{\parallel} = \la'(\xi), \:\:\: \Lambda_{\perp} = \la(\xi)/\xi. 
\end{equation}

\noindent
From Eqs. (\ref{eq:ode_xn}) and (\ref{eq:efa_xn}), equations for the coupling parameters are

\begin{gather}
    \frac{d \ecp_{\rho}}{d \tau} = \frac{\per}{2 \tau \ed_{\rho}} -  \left( \frac{1}{\tau_{\rho}} + \frac{1}{\tau} + \frac{1}{\ed_{\rho}} \frac{d \ed_{\rho}}{d \tau}\right)\ecp_{\rho}, \label{eq:ode_cp} \\
\tau_{\parallel} = \frac{\xi \la'(\xi)}{\la(\xi)}, \:\:\: \tpe = \frac{2 \la(\xi)}{\xi - \la(\xi)}.
\end{gather}

\noindent
Taking the second derivative in Eq. (\ref{eq:ode_difc}) and substituting Eq. (\ref{eq:ode_cp}) allows one to exclude coupling parameters from equations for diffusion coefficients:

\begin{equation}
    \tau^2_{\rho}\frac{d^2 \ed_{\rho}}{d \tau^2} + \tau_{\rho} \frac{d \ed_{\rho}}{d \tau} \left( 1 + \frac{2 \tau_{\rho}}{\tau} \right) + \ed_{\rho} \frac{\tau_{\rho}}{\tau} = \left( \Delta \difc_{\rho} + \difc \right) \frac{\tau_{\rho}}{\tau},  \label{eq:ode_difc2} 
\end{equation}

\noindent
where $\Delta \difc_{\rho} = \per^2 \Lambda_{\rho} \tau_{\rho}$.
The general solution can be written as 

\begin{equation}
    \ed_{\rho}(\tau) = \Delta \difc_{\rho}  + \difc + \left( C_1 + C_2 e^{-\tau/\tau_{\rho}} \right)\frac{\tau_{\rho}}{\tau},
\end{equation}

\noindent
where $C_1$ and $C_2$ are constants.
Demanding $\ed_{\rho}(0)$ to be finite, 
leads to $C_1 = - C_2 = \ed_{\rho}(0) - \Delta \difc_{\rho}  - \difc$, \textit{i.e.},

\begin{equation}
    \ed_{\rho}(\tau) =\Delta \difc_{\rho}  + \difc - \Big(\Delta \difc_{\rho} +\difc - \ed_\rho(0)\Big) \left( 1 - e^{-\tau/\tau_{\rho}} \right)\frac{\tau_{\rho}}{\tau}.
\end{equation}

\noindent
And then, with the initial conditions $\ed_\rho(0) = \difc$,
we get for the coupling parameters 

\begin{equation}
    \ecp_{\rho}(\tau) = \frac{\per}{2 \ed_{\rho}(\tau)} \left(1 - e^{-\tau/\tau_{\rho}}\right) \frac{\tau_{\rho}}{\tau}.
\end{equation}

\noindent
The local orientation vector $\dir$  within the EFA becomes

\begin{align}
    \mathcal{N}_i &= \bigg\{ \av{n_i}  + \ecp_{jk} \left( \av{n_i n_j} - \av{n_i} \av{n_j}\right) \left( x_k - \av{x_k}\right) \bigg\} P_{\text{EFA}} \notag \\
    &= \left\{ \la(\xi)h_i  + \ecp_{jk}\left(\frac{\la(\xi)}{\xi} \delta_{ij} + \left[\la_2(\xi) - \la^2(\xi)\right]h_i h_j \right)(x_k - \av{x_k})  \right \} P_{\text{EFA}}.
\end{align}

\end{document}